\documentclass[usenatbib]{mn2e}
\usepackage{graphicx}

\newcommand{\Msun}{\rm{\,M_\odot}}
\newcommand{\Mpch}{\rm{\,Mpc/h}}
\begin{document}
\title{Angular momentum and clustering properties of early dark matter halos}
\author[Davis \& Natarajan] {Andrew J. Davis,$^1$ and Priyamvada  Natarajan$^{1,2,3}$ \\
$^1$Department of Astronomy, Yale University, P.O. Box 208101, New Haven, CT 06520-8101, USA\\
$^2$Department of Physics, Yale University, P.O. Box 208120, New Haven, CT 06520-8120, USA\\
$^3$Radcliffe Institute for Advanced Study, Harvard University, 10 Garden Street, 
Cambridge, MA 02138, USA}

\maketitle
\begin{abstract}
In this paper we study the angular momentum properties of simulated
dark matter halos at high redshift that likely host the first stars in
the Universe. Calculating the spin distributions of these $10^6 -
10^7 \Msun$ halos in redshift slices from $z = 15 - 6$, we find that
they are well fit by a log-normal distribution as is found for lower
redshift and more massive halos in earlier work. We find that both the
mean value of the spin and dispersion are largely unchanged with
redshift for all halos. Our key result is that subsamples of low and
high spin $10^6 \Msun$ and $10^7 \Msun$ halos show difference in
clustering strength. In both mass bins, higher spin halos are more
strongly clustered in concordance with a tidal torquing picture for
the growth of angular momentum in dark matter halos in the CDM
paradigm.
\end{abstract}

\section{Introduction}

In the standard $\Lambda$CDM paradigm of structure formation, dark
matter over-densities accrete mass, forming deep potential wells into
which baryons condense and cool to eventually form stars. The details
of the cooling and star formation process are quite complex, and the
physics of the first episode of star formation and that of subsequent
generations is not completely understood. There have been detailed
numerical studies of the formation of the first stars which suggest
that the masses of these stars were skewed to the massive end
\citep{abel02,bcl02,omukai03,oshea07}. Numerical simulations are
challenging owing to the large dynamical range needed as well as the
complexities of hydrodynamical modeling required to capture the
essential physics. Current models of galaxy formation in the early
Universe are able to make some simplifying assumptions to make the
simulations of subsequent star formation easier \citep[see, e.g.,][]{grief08,johnson08,ricotti02a,ricotti02b,wise07,wise08}.
However, much work still needs to be done to fully uncover the details
of galaxy formation, including a better understanding of the
properties of the dark matter halos and how the dark matter properties
affect baryonic structure formation.
 
In numerical simulations, three basic properties of dark matter are
followed: mass, position, and velocity.  From these three properties,
much has been learned about the buildup and assembly of early dark
matter halos, including halo mass functions, galaxy clustering, halo
shapes, and halo spins \citep{reed07,jch01,reed03,reed08,maccio08}. Refinement
in our understanding of dark matter halos is needed to gain
insight into how the baryonic component of galaxies couples with the
dark matter (e.g., angular momentum transfer, stellar and AGN feedback). A key missing piece is the role of
angular momentum in the formation and modulation of structure
formation. This paper focuses on the angular momentum and clustering 
properties of dark matter halos in the early ($z =15 - 6$) Univers, 
roughly $2$ Gyr after the Big Bang in the concordance model. The goal 
of this work is to understand more about the early, low mass dark 
matter halos that likely host the earliest stars and galaxies in a 
statistically large sample.

\section{Origin of angular momentum in dark matter halos}

In the standard $\Lambda$CDM paradigm, angular momentum is acquired by
dark matter halos due to tidal torquing by their neighbors
\citep{hoyle49}. During the linear phases of proto-galactic evolution
the angular momentum grows to first order in proportion to time $t$.
\citet{peebles69} found from the study of collapsing spherical regions
that the growth in the angular momentum $\vec{J}$ occurs in the second
order only as $t^{5/3}$. In such regions growth occurs purely as a
result of convective effects on the bounding surface. This was
calculated using linear theory to find the growth rate of the angular
momentum within a co-moving spherical region of an expanding Einstein
de-Sitter Universe. \citet{doro70} showed that while the angular
momentum of a proto-galaxy did grow $\propto\,t$ for a flat Universe,
the Peebles result was a consequence of the imposed spherical symmetry
and was verified numerically by \citet{white84}. We briefly outline
the process of angular momentum acquisition by a dark matter halo in
what follows.

Consider the growth of fluctuations in an expanding Friedmann Universe
filled with pressureless matter with density, $\rho(\vec{r},t)$, and
let
\begin{eqnarray}
\delta(\vec{r},t) = \frac{\rho(\vec{r},t)}{\rho_0(\vec{r},t)}\,-1; 
\,\,\,\delta(\vec{x},t) = b(t)\,\delta_0(\vec{x}).
\end{eqnarray}
The local over-density can be written as a separable function of time
and a co-moving coordinate $\vec{x}$ related to $\vec{r}$ by $\vec{r}
= a(t)\vec{x}$, where $a(t)$ is the cosmological expansion factor. The
trajectory of a dust particle in linear theory is given by:
 \begin{eqnarray}
 \vec{r}(\vec{q},t) = a(t) \,\vec{x}(\vec{q},t) = a(t)\,\left[\vec{q} - b(t)\,\nabla \phi(\vec{q})\right]
 \end{eqnarray}
where $\vec{q}$ is the Lagrangian coordinate defined as the $\vec{x}$
position of the particle as $t \rightarrow 0$ and $\phi$ the peculiar
gravitational potential. If $\left<\delta^2\right> \ll 1$, a condition
that is satisfied over the period of time preceding the collapse only
when the initial density field has a coherence length of the
proto-galactic scale. In general, this will cease to describe the
detailed evolution of the matter distribution long before galaxies
form. The spin angular momentum of the material that makes up a
proto-galaxy can be written as,
\begin{eqnarray}
\vec{J}(t) = \rho_0\,a^5\,\int_{V_L}\,(\vec{x} - \mathbf{x})\,X\,\dot{\vec{x}}\,d^3q
\end{eqnarray}
Since the leading term in $\dot{\vec{x}}$ is first order and is
parallel to $\vec{x}$, the 2nd order term gives the non-zero
contribution to the above integral. Writing it explicitly as an
integral over the surface (using Gauss's theorem), we have:
\begin{eqnarray}
\vec{J}(t) = - a^2\,\dot{b}\,\epsilon_{ijk}\,T_{jl}\,I_{lk}
\end{eqnarray}
where $T$ is the tidal tensor and describes the local deformation at
$\vec{q}$ and $I$ is the inertia tensor of the matter in $V_{L}$. This
tensor product is used to calculate the tidal torque on an extended
body in the tidal field. Angular momentum is acquired as the first
order tidal field couples to the zeroth order quadrupole moment of the
irregular boundary of the proto-galaxy. The principal axes in general
do not coincide as $T$ defines the shape and disposition of the
neighboring perturbations and $I$ depends only on the shape of the
proto-galaxy. This outlines in principle the origin and offers insight
into the calculation of angular momentum of simulated dark matter
halos.

\section{Method}

We ran a series of N-body simulations to follow the growth of dark
matter halos from $z\approx 100$ down to $z=6$.  In order to study
halos that likely host Pop III stars, we chose the particle mass such
that a $10^6 \Msun$ dark matter halo would have $\sim$ 100 particles.
For $512^3$ particles, this requirement sets the box size at $2.46
\Mpch$ and the particle mass at $M_{DM} = 7.3 \times 10^3 \Msun$.  The
initial conditions were generated using a parallelized version of
Grafic \citep{mpgrafic}, which calculates the Gaussian random field
for the dark matter particles.  We used Gadget-2 \citep{springel05} to
follow dark matter particles down to a redshift of $z=6$, with output
snapshots at $z=15,12,11,10,9,8,7,6$. We used the WMAP3 \citep{wmap3}
cosmological parameters for the runs presented here, [$\Omega_M,
\Omega_\Lambda, \Omega_b, h, n, \sigma_8$] = [0.238, 0.762, 0.0416,
0.732, 0.958, 0.761].

For each snapshot, we used a friends-of-friends code to find
individual halos, using a linking length of $0.2$ times the initial
particle spacing. Once the halos were identified, each particle in
the halo was tested to see if it was actually bound to the halo. We
used SKID \citep{skid} to do the unbinding. SKID finds the potential
and kinetic energies for all particles in a given halo and removes
the most unbound particle from the halo. Successive iterations are
performed until all particles are either bound or there are no more
particles in the halo.  Without unbinding, there was excess power in both tails of the spin distribution.  

Once unbound particles were removed, for each halo that had more than
$50$ dark matter particles, we calculated the dimensionless spin
parameter,$$\lambda = \frac{J|E|^{1/2}}{GM^{5/2}},$$ where $J$ is the
total angular momentum, $E$ the total energy (kinetic plus potential),
and $M$ the total mass of the halo.  For the most massive halos, with
more than $10^4$ particles, we used the hierarchical force calculator
Treecode v1.4 \citep{treecode} to get the potential energy, inducing an error of $0.2\,\%$ in the potential energy as compared to direct summation.

\section{Results}

\begin{figure*}
\centering
\includegraphics[scale=0.6, angle = 90]{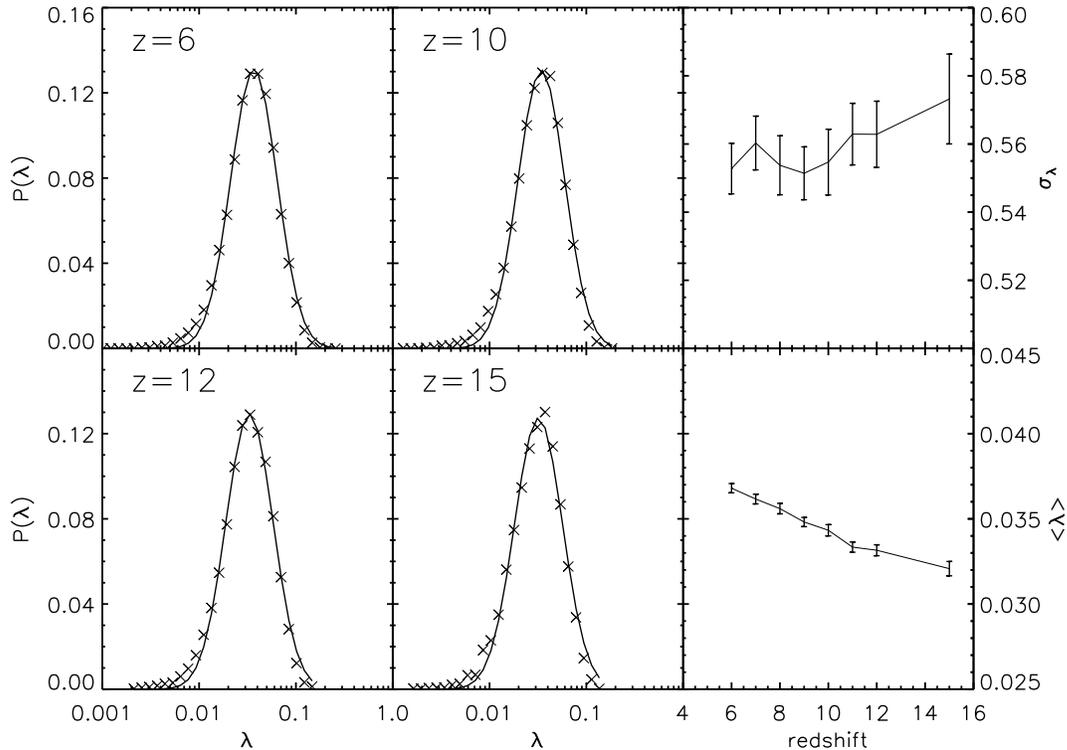}
\caption{Spin parameter distribution for all halos at four redshifts.
The solid curve is a log-normal fit to the data.  The right two panels
show the change in mean $\lambda$ and the standard deviation for the
log-normal fit for all redshift outputs used in this study.}
\label{lambda_tot}
\end{figure*}

Figure \ref{lambda_tot} shows the spin distribution at four redshifts
($z=6,10,12,15$).  As has been found for more massive
halos at low redshift  \citep[e.g.,]{bett07,bailin05,cole96, steinmetz95,
warren92}, the distribution is log-normal, with a slight
excess at low spin values. At $z=10$, \citet{jch01} also found a
log-normal distribution, with $\left<\lambda\right> = 0.033$ and
$\sigma = 0.52$, while we find $\left<\lambda\right> = 0.04$ and
$\sigma = 0.55$ at the same redshift.  These differences may be due to
the increased resolution, volume, and number of particles in our
study. In addition, \citet{jch01} used a different halo finding algorithm (SKID's density maxima finder), and found a Press-Schecter mass function.  However, our mass function matches the \citet{reed07} modification to the Sheth-Tormen mass function.   Finally, the differences between their results and our work may be due to different cosmological
parameters chosen, particularly $\sigma_8$.  In future work, we plan
to study the effect of mass resolution on the derived angular
momentum properties as well as the effect of changing $\sigma_8$.

The right column in Figure \ref{lambda_tot} shows the trend in the
mean and standard deviation of the fitted log-normal over redshift.  The error bars show the $1\sigma$ estimated errors on the mean and standard deviation of the fitted log-normal.
We found only slight variation in the mean spin, with $\lambda =
0.0378$ at $z=15$ increasing to $\lambda = 0.0426$ at $z=6$.  The
small increase in spin over time is most likely due to stronger tidal
torques from more massive halos that collapse at later times.  The
standard deviation decreases slightly from $\sigma = 0.573$ at $z=15$
to $\sigma = 0.553$ at $z=6$.

\begin{figure*}
\centering
\includegraphics[scale=.6, angle=90]{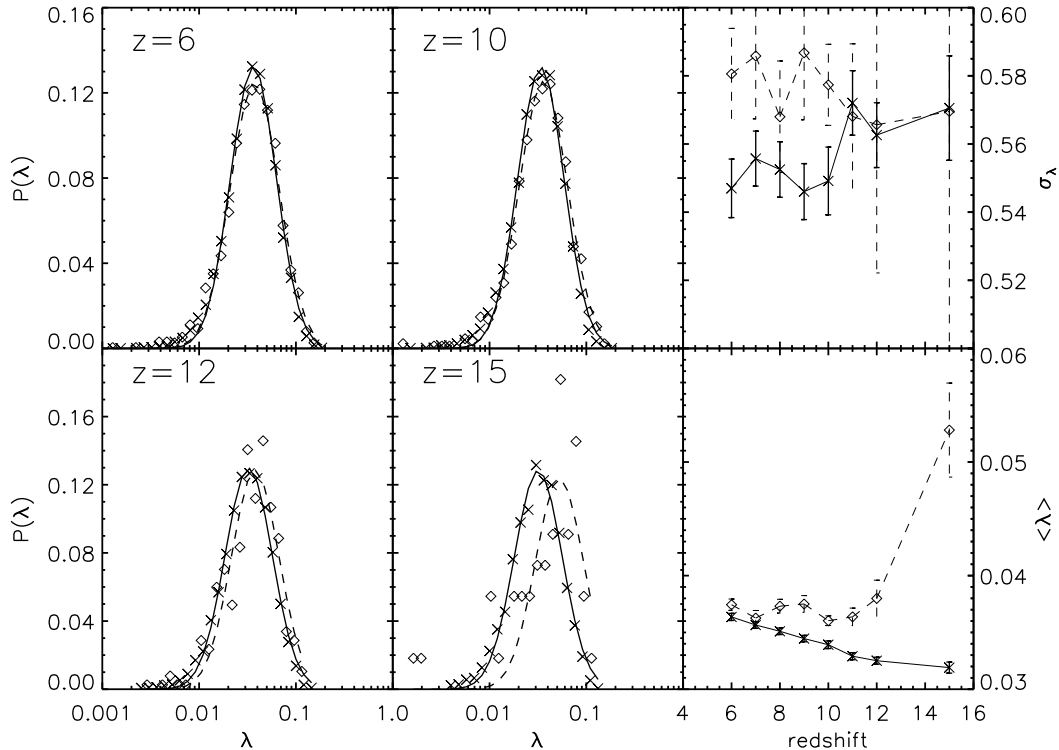}
\caption{Spin parameter distribution for two halo mass bins, $10^6
\Msun$ (crosses) and $10^7 \Msun$ (diamonds).  The solid ($10^6\Msun$)
and dashed ($10^7 \Msun$) curves are log-normal fits to the data, and
the right two panels show the change in mean $\lambda$ and the
standard deviation for the log-normal fit for each of the mass bins.}
\label{lambda_mbin}
\end{figure*}

We show in Figure \ref{lambda_mbin} the spin distribution for halos
with mass within $\pm 0.2$ dex of $10^7 \Msun$ and $10^6 \Msun$ at the
same redshifts as Figure \ref{lambda_tot}.  For the $10^7 \Msun$
halos, we find that $\left<\lambda\right>$ does not monotonically
increase with redshift as it did for the whole sample, and
$\left<\lambda\right>$ is systematically larger than the sample mean
at any given redshift.  There is also a sharp increase in
$\left<\lambda\right>$ at $z=15$, possibly because these halos were
still in the process of forming and may have significant infalling
mass clumps.  However, there are only 55 halos in this mass bin at $z=15$, making it difficult to reliably fit the sample with a log-normal curve.  
The $10^6 \Msun$ halos show spin distributions more
similar to the entire sample, with a slowly increasing mean spin from
$z=15$ to $z=6$.

\section{Correlation function of high-redshift dark matter halos}

We also studied the clustering properties of dark matter halos selected
by their angular momentum. It is well known that more massive halos
are strongly clustered \citep[e.g.,][]{mo02,st99,lemson99}.
\citet{lemson99} found no correlation between $\lambda$ and
environment for halos with $M_{tot} > 10^{12} \Msun$ at $z=0$,
\citep[see also][]{cs08,maccio07,berta08,bett07}.  This diagnostic is potentially
important as a halo of mass $M$ with larger angular momentum may take
longer to collapse and therefore longer to form stars than a lower
spin halo of the same mass.  Thus, low spin halos could preferentially
host star formation earlier.  This in turn may allow feedback
processes such as ionizing photons, stellar winds, and supernovae to
become important earlier than in their high spin counterparts. These
are important consequences for galaxy formation, and the detailed role
of the spin of the halo on the collapse of baryons will be explored in
a subsequent paper.

\begin{figure*}
\begin{centering}
\includegraphics[width=0.6\textwidth, angle=90]{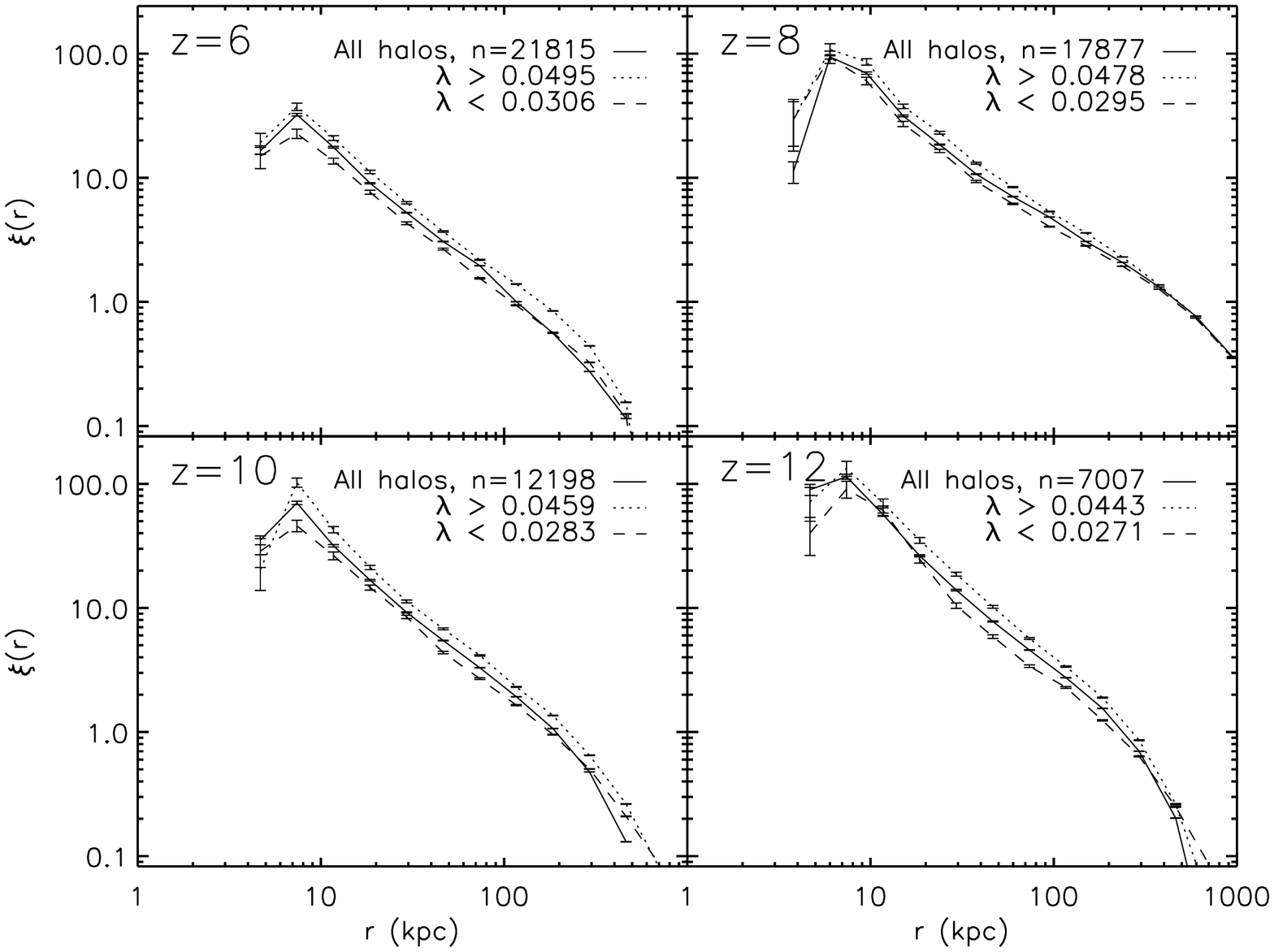}
\includegraphics[width=0.6\textwidth, angle=90]{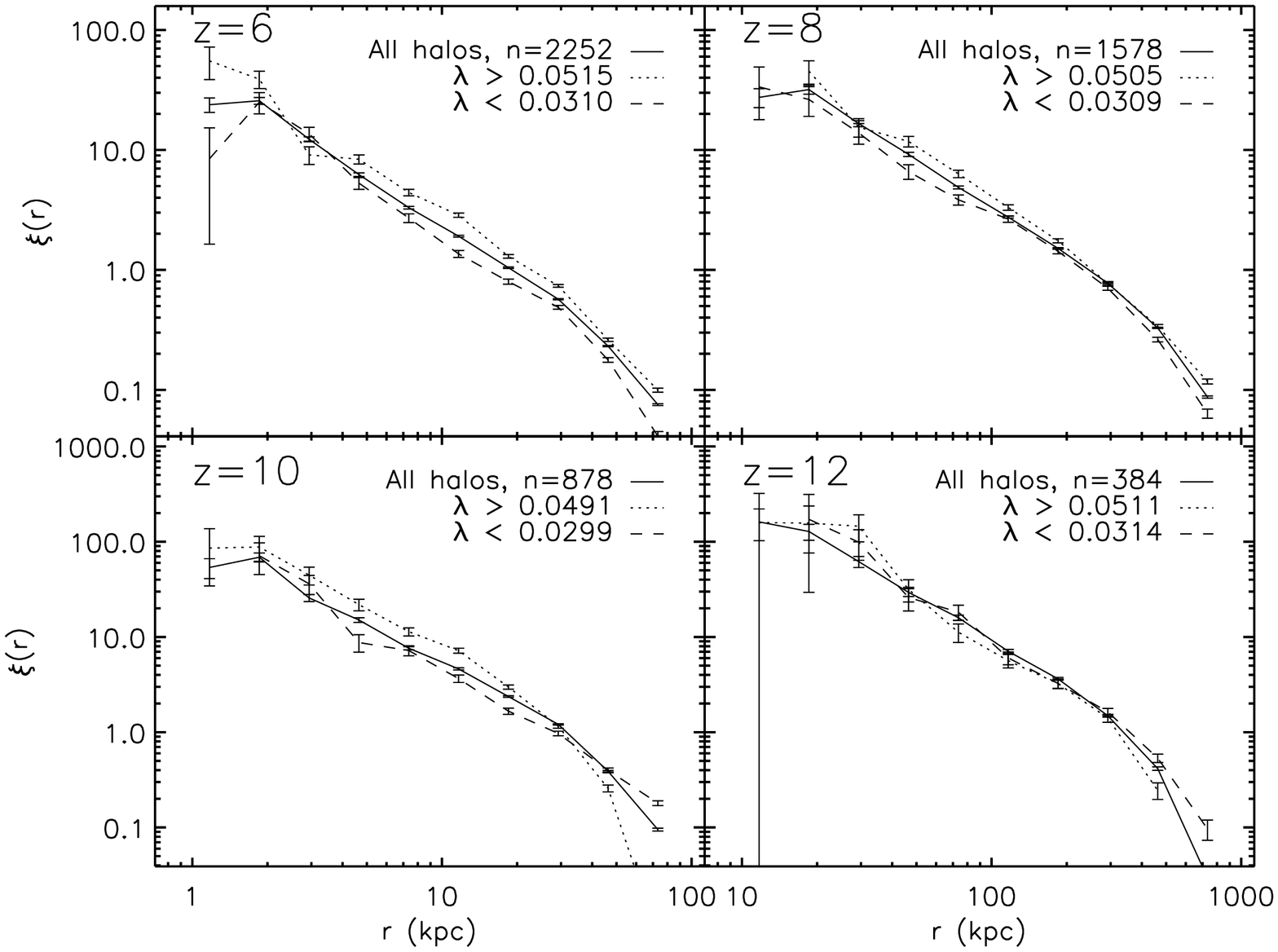}
\caption{Correlation function for $10^6 \Msun$ (top panel) and $10^7
\Msun$ (bottom panel) halos as a function of comoving separation, $r$.
Each plot shows the correlation function for all halos in the mass bin
(solid), as well as for a cut based on the halo spin parameter.  The
cuts were chosen so that one third of the halos would fall in the high
(dotted) and low (dashed) bins.}
\label{corr}
\end{centering}
\end{figure*}

We calculate the two-point correlation function, $\xi(r)$, using the standard definition,
\begin{equation}
\xi(r) = \frac{DD(r) RR(r)}{RD(r)^2} -1,
\end{equation}
where $DD(r)$ is the number of halo-halo pairs within a separation of
$r-dr/2$ and $r+dr/2$, $RR(r)$ the same for a random distribution, and
$RD(r)$ is the cross-correlation between the data set and the random
distribution. The quantity $\xi(r)$ represents the excess probability
of finding a halo at distance $r$ compared to a random distribution.

Figure \ref{corr} shows the correlation function, $\xi(r)$, for halos
grouped by spin parameter for a given mass.  The top panel is for
halos with mass $10^{6} \Msun$, and the bottom panel for $10^{7} \Msun$
halos.  We separated the halos out by mass to delineate the effect of
spin from that of mass. We first computed the overall correlation
function for all halos in the given mass range, and then used two
sub-samples based on their spin parameter. The cuts in spin parameter
were chosen so that one third of the halos were in the high and low
spin bins respectively.  The error bars show Poisson errors in $DD(r)$,
$$\sigma_\xi = \frac{RR(r)}{RD(r)^2} \sigma_{DD(r)}.$$
When counting halo pairs for the spin cut
samples, we employed two methods. First, we counted only pairs where
both halos were in the given bin, and these results are shown in
Figure \ref{corr}.  We also counted the pairs where only one halo was in
the high or low spin bin. This method of counting produced similar
results as the first method.

We find that there is a distinction in the correlation function when
the data are separated by the spin parameter. High spin halos are more
strongly clustered than low spin halos, similar to the result found by
\citet{bett07} in the Millennium Simulation. We also find that the difference in clustering is preserved over a range of redshifts.  The
excess clustering is due to the fact that in a denser environment,
halos feel stronger tidal torques, and thus have larger angular
momentum and spin parameters.  The correlation function can be fit to
a power-law, $\xi(r) = (r/R)^\gamma$, where $R$ is the correlation
length.  We find the high spin halos have a correlation length on
average $25\%$ larger than the low spin halos over the various mass
and redshift slices used here.  The slope, $\gamma$, did not show any correlation with spin, mass, or redshift.

\section{Discussion}

In this paper, we have examined the angular momentum distribution of
simulated high redshift dark matter halos and their correlation
functions. While the angular momentum distributions for this
population are well fit by log-normal distributions as is the case for
lower redshift and more massive halos, we find that the clustering
properties of the high redshift population depend on the value of
spin. For a given mass bin selecting by spin, we find that the
correlation function is higher for high spin halos. This is an
important trend and appears to be robust. Although in this study we
have restricted ourselves to the analysis of dark matter only
simulations, the different correlation lengths of high and low spin
halos are likely to have an important impact in feedback from these
early galaxies. Pop III stars appear to have a large impact on their
environment due to radiative \citep{johnson07,whalen08b} and supernova
\citep{grief07,whalen08} feedback. 

We note here that in a recent paper
\citet{oshea07} tracked the spin parameters of halos that formed the
first stars in 12 cosmological random realizations and did not find a
correlation between halo spin and collapse time. The resolution and
analysis pursued by them differ from ours as they zoom in and track
(by explicit selection) the properties of the dark matter halo that
hosts the first star.  Our analysis is distinct by construction as we
are measuring the spin parameters of an ensemble of dark matter halos
and arguing that the higher spin halos might in general commence the
collapse and formation of its first star later. As shown in Figure
\ref{lambda_mbin}, the distribution of halo spin for the high mass bin
at $z=15$ is shifted to higher spins when compared to the general
population. Thus, the larger halos that would host Pop III stars will
have larger spins, but the largest spin that \citet{oshea07} followed
was for $\lambda \approx 0.1$, which is at the half-max of the log-normal fit.  
One halo with this spin may not be representative of all high spin, massive halos, which may show a delayed baryonic collapse.  While this
is beyond the scope of this paper, we intend to pursue the
consequences of our results for baryonic collapse.

Our results are in agreement with what \citet{bett07} found for the correlation of 
dark matter halos at $z = 0$ in the Millenium Run. \citet{berta08}  found trends in the  
SDSS spectroscopic sample of local galaxies relating the estimated dark matter spin parameter with the galaxy's baryonic
properties, such as stellar mass and recent star formation
histories.  Thus, angular momentum may be a key to determining the epoch of
formation, size, and structure of the stellar and gaseous disks in
galaxies. In the early Universe however, it remains unclear how much
effect angular momentum will have on the first generation of galaxy
formation, although our results are suggestive.

\section*{Acknowledgments}

The authors are grateful for the prompt and thoughtful comments of the referee, and for time allocation on two of Yale's super-computing
clusters, Bulldog A and Bulldog H.  A. J. D. is grateful for many helpful discussions with Jason Tumlinson and Kyoung-Soo Lee.

\end{document}